\documentclass[a4,10pt]{article}
\usepackage{amssymb}
\usepackage{amsmath}
\usepackage{latexsym}
\usepackage{array}
\usepackage{bm}
\usepackage{theorem}
\usepackage{graphicx}
\usepackage{exscale}

\theoremstyle{break}
\newtheorem{Theorem}{Theorem}

\def\Proof{\hfil\break{\bf Proof}\;\;\;\;}
\def\Proofprop1{\hfil\break{\bf Proof of proposition \ref{Prop1}}\;\;\;\;}
\newtheorem{Proposition}{Proposition}
\newtheorem{Lemma}{Lemma}
\newtheorem{Corollary}{Corollary}
\newtheorem{Definition}{Definition}

\def\hbreak{\vspace*{2mm}\hfill\break\noindent}

\def\Z{\mathbb{Z}}

\def\C{\mathbb{C}}

\def\qed{\hfill\hbox{\rule[-2pt]{3pt}{6pt}}}

\begin{document}

\title{Algebraic entropy of an extended Hietarinta-Viallet equation}
\author{Masataka Kanki$^1$, Takafumi Mase$^1$ and Tetsuji Tokihiro$^1$\\
\small $^1$ Graduate School of Mathematical Sciences,\\
\small University of Tokyo, 3-8-1 Komaba, Tokyo 153-8914, Japan
}
\date{}
\maketitle 

\begin{abstract}
We introduce a series of discrete mappings, which is considered to be an extension of the Hietarinta-Viallet mapping
with one parameter. We obtain the algebraic entropy for this mapping
by obtaining the recurrence relation for the degrees of the iterated mapping.
For some parameter values the mapping has a confined singularity, in which case the mapping is equivalent to a recurrence relation between six irreducible polynomials.
For other parameter values, the mapping does not pass the singularity confinement test.
The properties of irreducibility and co-primeness of the terms play crucial roles in the discussion.

\texttt{MSC: 37K10, 39A20}

\texttt{Keywords: Hietarinta-Viallet map, algebraic entropy, integrability criterion, singularity confinement}
\end{abstract}

\section{Introduction}
Singularity confinement test (SC test) is one of the most famous integrability criteria for discrete equations \cite{SC}.
It is introduced as a discrete analogue of the Painlev\'{e} test \cite{Conte}. The Painlev\'{e} test determines whether the given ordinary differential equation possesses movable singularities.
The absence of movable singularities well predicts the integrability of the continuous equation.
Analogously, according to the SC test, the discrete equation is integrable,
if the spontaneously appearing singularities disappear after a finite iteration steps.
As we shall describe later with our main target (the Hietarinta-Viallet equation and its extension), the SC test is not equivalent to the integrability of some discrete equations.
We have another test for integrability: zero algebraic entropy criterion.
The algebraic entropy estimates the increasing rate of the degrees of the iterated mapping \cite{BV}.
Let $\phi$ be a recurrence relation for a sequence $\{x_n\}_{n=0}^{\infty}$, which determines $x_{n+1}$ as a rational function $F(x_{n},x_{n-1},\cdots)$. Let us suppose that the degree of $F$ is $d$.
Let us denote the degree of the iterated mapping $\phi^n=\underbrace{\phi\circ \phi\circ \cdots \circ\phi}_n$ as $d_n$.
The {\em algebraic entropy} $\lambda_{\phi}$ is defined as
\[
\lambda_{\phi}:=\lim_{n \to \infty}\frac{\ln d_n}{n},
\]
which is always convergent to a non-negative real value.
The {\em dynamical degree} of the mapping $\phi$ is defined as $\lim_{n\to \infty} (d_n)^{1/n}$, and is equal to $e^{\lambda_{\phi}}$.
The criterion states that the integrability of the mapping $\phi$ is closely related to the fact that $\lambda_{\phi}=0$.
Our understanding is that, in most cases, $\lambda_{\phi}=0$ if and only if $\phi$ is integrable.
In some cases, however, the results from the SC test and the zero algebraic entropy test conflict with each other.
As for the extended Hietarinta-Viallet equation we shall deal with, the result depends on the parity of a parameter introduced in the equation.
One of the ways to obtain the algebraic entropy is to construct a recurrence relation for $d_n$. Diller and Favre proved that
there exist a finite order recurrence for $d_n$, if the mapping $\phi$ is a birational mapping over $\mathbb{P}^2$ \cite{DF}.
Note that in their work, the degrees are counted for the homogeneous representation in $\mathbb{P}^2$, while, in our paper, we mainly use the degrees over $\mathbb{P}^1\times \mathbb{P}^1$. This difference does not affect the value of the algebraic entropy.
The Hietarinta-Viallet equation \cite{HV} passes the singularity confinement test, but has chaotic solutions, whose existence is an indication of the non-integrable nature of the equation.

We consider the following extension of the Hietarinta-Viallet mapping:
\begin{equation}
x_{n+1}=-x_{n-1}+x_n+\frac{1}{x_n^k} \qquad (k=2,3,4,...),
\label{HVeq}
\end{equation}
and obtain the algebraic entropy of the equation \eqref{HVeq}.
Let us denote the algebraic entropy of the mapping \eqref{HVeq} as $\lambda_k$. The same extension for even $k$ is studied in \cite{MW} in terms of full deautonomisation method, and the value of $\lambda_k$ is conjectured for even $k\ge 2$, perfectly agreeing with our result here.
Let us comment that the coefficient $1$ in the term $1/x_n^k$ is not essential: by a gauge transformation $y_n=c^{\frac{1}{k+1}} x_n$, equation \eqref{HVeq} is equivalent to
$y_{n+1}=-y_{n-1}+y_n+c/y_n^k$, which has exactly the same algebraic entropy as \eqref{HVeq}.
A non-autonomous extension is found in \cite{MW}.
The original Hietarinta-Viallet equation \cite{HV} is recovered when $k=2$.
It has been conjectured in \cite{HV} that $\lambda_2=\ln (3+\sqrt{5})/2=0.962\dots$,
and has been proved in \cite{BV} by constructing the recurrence relation for the degrees of the iterated mappings.
Takenawa obtained the algebraic entropy of the Hietarinta-Viallet mapping through a geometric description of the space of initial conditions \cite{Takenawa,Takenawa2}.
The evolution of the equation induces an action on the Picard group generated by the exceptional divisors introduced in order to realize the mapping as an automorphism over a rational surface.
The action on the Picard group is expressed as a matrix. The largest eigenvalue of this matrix gives the dynamical degree, the logarithm of which is the algebraic entropy.

Our main results are corollaries \ref{evenentropy} and \ref{algodd}, which give the algebraic entropy $\lambda_k$ for even $k$ and odd $k$ separately: $\lambda_k=\log((k+1+\sqrt{(k-1)(k+3)})/2)$ for even $k\ge 2$, and $\lambda_k=\log ((k+\sqrt{k(k+4)})/2)$ for odd $k\ge 3$.
For example, we have $\lambda_3=\ln (3+\sqrt{21})/2=1.332\dots$ and $\lambda_4=\ln (5+\sqrt{21})/2=1.566\dots$.
The main reason for the difference between the case of even $k$ and the odd one is the singularity structure of the mapping \eqref{HVeq}.
The mapping \eqref{HVeq} passes the singularity confinement test for $k=1$ and even $k=2,4,6,\cdots$. However, for odd $k=3,5,7,\cdots$, it does not pass the SC test.
When the mapping passes the singularity confinement test as it does for even $k$ here, it should be possible to conjugate the map to an automorphism by  blowing-up the domain of definition $\mathbb{P}^1\times \mathbb{P}^1$ at the singularities of the mapping, just like Takenawa has done 14-times blowing-ups for $k=2$ in \cite{Takenawa,Takenawa2}. However, the number of blowing-ups needed is not readily obtained and could be quite large for even $k\ge 4$.
Moreover, as for the odd $k$ case, the construction of the space of initial condition is expected to be impossible because of the non-confining property (We will come back to this topic in the conclusion). Therefore we do not take this geometric approach and use an algebraic, and rather an elementary method, by investigating the factorizations of the iterates.
The factorization for the iterates of the Hietarinta-Viallet mapping ($k=2$) is observed in \cite{Viallet}. Our results are related to \cite{Viallet} and also include generalized results and rigorous proofs.
The exact form of the factorization of the general term into some irreducible polynomials tells us the recurrence relation for the degrees of the iterated mappings. The largest real root of the characteristic polynomial of this recurrence relation gives the dynamical degree.
To obtain the factorization forms, the irreducibility of each factor plays an important role.
The algebraic entropy is immediate from the recurrence relation as in \cite{BV}.

At the last section of this paper in theorem \ref{evenirredthm}, we prove the {\em irreducibility} of the terms of the mapping \eqref{HVeq} for even $k$, by refining a lemma \ref{lem5} used to obtain the algebraic entropy.
The {\em irreducibility and co-primeness} are conjectured to be deeply related to the singularity structure and the integrability of the given discrete mappings, from our previous results that they are equivalent to the integrability in the case of the discrete KdV equation, one type of mappings related to the Somos-$4$ sequence, and the discrete Toda equation \cite{dKdVSC2,dtoda}.
Our investigation of the algebraic entropy in terms of the irreducibility and co-primeness in this paper is expected to be applicable to other integrable and non-integrable discrete equations, such as the {\em linearizable} type QRT mappings \cite{Ramani}, and non-QRT type mappings \cite{Tsuda}.

\section{Algebraic entropy of the mapping \eqref{HVeq}}
Let us define the mapping \eqref{HVeq} over the projective space $\C\mathbb{P}^2$, and write the evolution using the homogeneous coordinate $[p_n:q_n:r_n]=[x_{n}:x_{n-1}:1]$. In the homogeneous coordinates, the point itself is unchanged by multiplying all the three variables by a common factor: i.e., $[P:Q:R]=[fP:fQ:fR]$ for $f\neq 0$. Then we have
\begin{subequations}
\begin{align}
p_{n+1}&=p_n^{k+1}-q_np_n^k+r_n^{k+1}\label{cp2_a},\\
q_{n+1}&=p_n^{k+1}\label{cp2_b},\\
r_{n+1}&=r_np_n^k\label{cp2_c}.
\end{align}
\end{subequations}
Note that we do not assume a minimal form for the homogeneous coordinates: i.e., we allow an existence of common factors among $p_n,q_n,r_n$.
We take the initial values as $p_0=a,\,q_0=b,\,r_0=c$.
Note that $x_{-1}=b/c$ and $x_0=a/c$.
Repeating equations \eqref{cp2_a} -- \eqref{cp2_b}, we obtain
\begin{align}
p_{n+1}&=p_n^k(p_n-p_{n-1}^{k+1})+c^{k+1}\left(p_{n-1}p_{n-2}...p_1p_0\right)^{k(k+1)}  \quad (n \ge 1) \label{eqp1}\\
&=p_n^k\left\{ -p_{n-1}^{k}p_{n-2}^{k+1}+c^{k+1}(p_{n-2}p_{n-3}...p_0)^{k(k+1)}\right\} \notag\\
&\qquad \qquad \qquad +c^{k+1}\left(p_{n-1}p_{n-2}...p_1p_0\right)^{k(k+1)}  \quad (n \ge 2). \label{eqp2}
\end{align}
For example the first three iterates of $p_n$ are as follows.
\begin{align}
p_1&=a^{k+1}+c^{k+1}-a^kb, \label{p1form}\\
p_2&=(c^{k+1}-a^kb)p_1^k+c^{k+1}a^{k(k+1)}, \label{p2form}\\
p_3&=\left\{(ca^k)^{k+1}-p_1^ka^{k+1} \right\}p_2^k+(ca^k)^{k+1}p_1^{k(k+1)}\notag\\
&=a^{k+1} \left\{(c^{k+1}a^{k^2-1}-p_1^k)p_2^k+c^{k+1}a^{k^2-1}p_1^{k(k+1)}  \right\}.\label{p3form}
\end{align}
Before going into the details, let us prepare two lemmas.
\begin{Lemma}\label{lem0}
We have
\begin{subequations}
\begin{align}
\deg (f+g)&\ge\left|\deg f - \deg g\right|, \label{degree1}\\
\deg (f+f^{-k})&=(k+1)\deg f, \label{degree2}
\end{align}
\end{subequations}
for any non-zero rational functions $f,g$, where the degree $\deg f$ of the rational function $f$ is defined as the maximum of the degrees of its constituent polynomials.
\end{Lemma}
\Proof
Let us express $f=f_1/f_2$ and $g=g_1/g_2$ with their constituent polynomials, where
$f_1$ and $f_2$ are coprime with each other (and the same for $g_1$ and $g_2$).
Let us take the greatest common divisor (GCD) of $f_1$ and $g_1$ as $h_1$, and the GCD of $f_2$ and $g_2$ as $h_2$:
\[
f_1=h_1 f_1',\, g_1=h_1 g_1',\quad f_2= h_2 f_2',\, g_2=h_2 g_2',
\]
where $f_i', g_i'$ $i=1,2$ are some polynomials.
Then the polynomial $f_2'$ should be coprime with $g_2',f_1',h_1$.
We also have that $g_2'$ is coprime with $f_2',g_1',h_1$.
From
\[
f+g=\frac{h_1(f_1'g_2'+g_1'f_2')}{h_2f_2'g_2'},
\]
we have
\[
\deg (f+g) \ge \deg (f_2'g_2'),
\]
since $f_2'g_2'$ does not factorize with the numerator.
We also have
\[
\deg (f_2'g_2')\ge \deg (f_2')=\deg (f)-\deg (h_2)\ge \deg (f)-\deg (g)
\]
Since the discussion is symmetric with $f$ and $g$
we have proved equation \eqref{degree1}. Next we compute
\[
f+f^{-k}=\frac{f_1^{k+1}+f_2^{k+1}}{f_2f_1^{k}}.
\]
Since $f_1$ and $f_2$ are coprime, the denominator and the numerator
do not share a factor. Thus equation \eqref{degree2} is proved.
\qed
\begin{Lemma}\label{lem1}
Let us suppose that $x_{-1}=0,\,x_0=a$ in \eqref{HVeq}.
Then $x_n$ is not identically zero as a rational function of $a$.
\end{Lemma}
\Proof
In the case of mapping \eqref{HVeq}, we have $\deg x_0=1$, $\deg x_1=k+1$. It is enough to show that $\deg x_n\ge 1$ for any positive integer $n$.
Let us prove $\deg (x_n)>\deg (x_{n-1})$ by induction.
Suppose that $\deg (x_n)>\deg (x_{n-1})$.
Since $\deg (x_n+x_n^{-k})=(k+1)\deg(x_n)$ from equation \eqref{degree2}, we have $\deg (x_{n+1})=\deg (-x_{n-1}+x_n+x_n^{-k}) \ge (k+1)\deg (x_n)-\deg (x_{n-1})>k\deg (x_n)>\deg (x_n)$, where we have used \eqref{degree1} in the first inequality.
Therefore $x_n$ cannot be identically zero.
\qed

We have that the algebraic entropy $\lambda_k$ of the mapping \eqref{HVeq} satisfies
\[
\lambda_k\ge \ln k,
\]
because we have $\deg x_n\ge k^{n-1}$ from the proof of lemma \ref{lem1}.
Therefore the extended Hietarinta-Viallet mapping \eqref{HVeq} has a positive algebraic entropy and is not supposed to be integrable.
However, the singularity structure deeply depends on the parity of the integer parameter $k\ge 2$. The mapping passes the singularity confinement test for even $k\ge 2$ (and for $k=1$), while in the case of odd $k\ge 3$ it does not pass the test.
We define a sequence $\{\beta_n\}_{n\ge 0}$ of polynomials of $k$, which will be proved in proposition \ref{Prop1} to be equal to the order ord$_a (p_n)$ of the factor {\em a} in $p_n$. The following formulae for $\beta_n$ are obtained recursively from equations \eqref{eqp1} and \eqref{eqp2} and the discussions in the next two subsections.
\begin{Definition}\label{defbeta}
\begin{itemize}
\item For even $k\ge 2$:
Let us define a sequence $\beta_n$ $(n\ge 0)$ by
$\beta_0=1,\,\beta_1=\beta_2=0,\, \beta_3=k+1$ and
$\beta_n:=  k(k+2)(k+1)^{n-4}$ for $n\ge 4$.

\item For odd $k\ge 3$:
Let us define a sequence $\beta_n$ $(n\ge 0)$ by
$\beta_0=1,\,\beta_1=\beta_2=0$ and
$\beta_n:=  k(\beta_{n-1}+\beta_{n-2})+(k+1)\beta_{n-3}$
for $n\ge 3$.
\end{itemize}
\end{Definition}
\begin{Definition}
We define a sequence of Laurent polynomials $\tilde{p}_n$ by
$\tilde{p}_n:=a^{-\beta_n}p_n$.
\end{Definition}
\begin{Proposition}\label{Prop1}
We have ord$_{a}(p_n)=\beta_n$ for all $k\ge 2$ and $n\ge 0$. In other words,
the function
$\tilde{p}_n(a,b,c) \in\mathbb{Z}[a,b,c]$ is a polynomial. 
Also we have $\tilde{p}_n(0,b,c) \neq 0$: i.e., $\tilde{p}_n$ does not have $a$ as a factor. 
\end{Proposition}
Proposition \ref{Prop1} states that the iterate $p_n$ is divisible by the factor $a$ exactly $\beta_n$ times.
Proof of this proposition depends on the parity of $k$, which will be treated in the following subsections separately.
\begin{Definition}
We define a new sequence $\{\alpha_n\}$ by
$\alpha_1:=0$ and
\begin{equation}
\alpha_n:=\beta_n-(\beta_{n-1}\alpha_1+\beta_{n-2}\alpha_2+...+\beta_1\alpha_{n-1}),
\label{alpha}
\end{equation}
for $n\ge 2$.
\end{Definition}
\begin{Definition}
We define an operator $T$ acting on the field of rational functions $\mathbb{C}(a,b,c)$ as substituting $(p_1=a^{k+1}+c^{k+1}-a^k b,q_1=a^{k+1},r_1=a^k c)$ in the variables $(a,b,c)$: i.e.,
for a rational function $f(a,b,c)$, we have
\[
(Tf)(a,b,c):=f(p_1,q_1,r_1).
\]
We define a sequence of new rational functions $\{p'_n\}$
with
\begin{equation}\label{pdash}
 p_n':=a^{-\alpha_n}\left( T p_{n-1}'\right),
\end{equation}
for $n\ge 1$ where $p_0':=a$.
\end{Definition}
The first four iterates are calculated as
$p'_1=p_1=\tilde{p}_1$, $p'_2=p_2=\tilde{p}_2$, $p'_3=a^{-(k+1)}p_3=\tilde{p}_3$,
$p'_4=a^{-\alpha_4}T(p'_3)=\cdots=a^{-\beta_4}p_1^{-(k+1)}p_4=p_1^{-(k+1)}\tilde{p}_4$, where $\beta_4=k(k+2)$ for even $k\ge 2$, and $\beta_4=k(k+1)$ for odd $k\ge 3$.
\begin{Lemma}\label{lem2}
We have the following three properties for $p'_n$ $(n\ge 1)$:
\begin{itemize}
\item $p'_n\in\mathbb{Z}[a,b,c]$,
\item $p'_n$ is not divisible by `$a$' in $\mathbb{Z}[a,b,c]$,
\item $p'_n$ satisfies the following relation
\begin{equation}\label{pnexp}
p_n=(p'_0)^{\beta_n}(p'_1)^{\beta_{n-1}}...(p'_n)^{\beta_0}.
\end{equation}
\end{itemize}
\end{Lemma}
\Proof
The proof is by induction. If $n=1$, the statements are satisfied because $p_1=p'_1=(p'_0)^{\beta_1}(p'_1)^{\beta_0}$.
Let us assume that
\[
p_{n-1}=(p_0')^{\beta_{n-1}}(p_1')^{\beta_{n-2}}\cdots (p_{n-1}')^{\beta_0},
\]
and assume that $p_1',...,p_{n-1}'$ are polynomials, none of which has a factor $a$.
By applying $T$ to both sides,
\begin{align*}
p_n&=(T(p_0'))^{\beta_{n-1}}(T(p_1'))^{\beta_{n-2}}\cdots (T (p_{n-2}'))^{\beta_1} T(p'_{n-1})\\
 &=a^{\sum_{j=1}^{n-1}\alpha_j \beta_{n-j}}(p_1')^{\beta_{n-1}}(p_2')^{\beta_{n-2}}\cdots (p_{n-1}')^{\beta_1} T(p'_{n-1}),
\end{align*}
where we have used the relation $T(p'_{m-1})=a^{\alpha_m}p'_m$ for $m=0, 1,\cdots, m-2$ in the second equality.
By using the definition of $\alpha_n:=\beta_n-\sum_{j=1}^{n-1}\beta_{n-j}\alpha_j$,
and by dividing the both sides by $a^{\beta_n}$ we obtain
\[
\tilde{p}_n=p'_n \cdot (p_1')^{\beta_{n-1}}\cdots (p_{n-1}')^{\beta_1}.
\]
Here we have used the relations $p_n=a^{\beta_n} \tilde{p}_n$ and $p'_n=a^{-\alpha_n}T(p'_{n-1})$.
Since none of the terms $\tilde{p}_n, p_1',...,p_{n-1}'$ has a factor `$a$' from proposition \ref{Prop1} and the induction hypothesis,
we have ord$_a(p'_n)=0$, which indicates that $p'_n$ is a polynomial and that $p'_n$ is not divisible by `$a$'. The relation
$p_n=(p'_0)^{\beta_n}(p'_1)^{\beta_{n-1}}...(p'_n)^{\beta_0}$
follows from $p'_0=a$ and $\beta_0=1$.
\qed

\begin{Lemma}\label{lem4}
The polynomial $p_n'$ is not divisible by a factor `$c$'.
\end{Lemma}
\Proof
None of $p_0'=1,\,p_1'=1-b,\,p_2'=-b(1-b)^k$ is $0$ for $a=1$ and $c=0$.
The equation \eqref{eqp2} tells us that for $c=0$, $p_{n+1}=-p_n^kp_{n-1}^kp_{n-2}^{k+1} \quad (n \ge 2)$.
Therefore $p_n\neq 0$ for all $n$ when $c=0$, which proves the lemma.
\qed

From here on we investigate the case of even $k$ and odd $k$ in separate subsections.
\subsection{The case of even $\boldsymbol k\ge 2$}
First let us prove the proposition \ref{Prop1} for even $k$.
\Proofprop1
The case of $n=0,1,2$ is trivial from expressions \eqref{p1form}--\eqref{p3form}. Note that we have $\tilde{p}_0=1$, $\tilde{p}_1=p_1$, $\tilde{p}_2=p_2$ and that $p_1(0,b,c)=c^{k+1}$, $p_2(0,b,c)=c^{(k+1)^2}$.
In the case of $n=3$,
\begin{equation}\label{p3d}
\tilde{p}_3=a^{k^2-1}(c^{k+1}p_2^k+c^{k+1}p_1^{k(k+1)})-p_1^kp_2^k,
\end{equation}
since $p_3=a^{k+1}\tilde{p}_3$.
We have $\tilde{p}_3(0,b,c)=-c^{k(k+1)(k+2)}$.

In the case of $n=4$, we have
\begin{align}
p_4&=p_1^{k+1}\left[ \left\{(ca^k)^{k+1}(p_1)^{k^2-1}-p_2^k   \right\}p_3^k+(ca^k)^{k+1}p_1^{k^2-1}p_2^{k(k+1)}  \right] \notag \\
&=a^{k(k+1)}p_1^{k+1}\left[ \left\{ c^{k+1}a^{k(k+1)}p_1^{k^2-1}-p_2^k   \right\}(\tilde{p}_3)^k+c^{k+1}p_1^{k^2-1}p_2^{k(k+1)}   \right]. \label{p4keisan}
\end{align}
Let us extract the last two terms without factor `$a$' in the parentheses $[ \ ]$ and deform them:
\[
-p_2^k(\tilde{p}_3)^k+c^{k+1}p_1^{k^2-1}p_2^{k(k+1)} 
=p_2^k \left[c^{k+1}p_1^{k^2-1}p_2^{k^2}-(\tilde{p}_3)^k  \right].
\]
From equation \eqref{p3d}, we have
\[
(\tilde{p}_3)^k=X a^{k^2-1}+(-1)^kp_1^{k^2}p_2^{k^2},
\]
where $X\in\mathbb{Z}[a,b,c]$ is some polynomial.
Thus we have
\begin{align*}
&p_2^k\left[ -(\tilde{p}_3)^k +c^{k+1}p_1^{k^2-1}p_2^{k^2}\right] =p_2^k\left[-a^{k^2-1}X+p_1^{k^2-1}p_2^{k^2}\left\{ c^{k+1}+(-1)^{k+1}p_1  \right\}\right] \\
&=a^k p_2^k \left[-a^{k^2-k-1}X+p_1^{k^2-1}p_2^{k^2}(b-a) \right],
\end{align*}
since $k$ is an even integer and $p_1=a^{k+1}+c^{k+1}-a^k b$.
Substituting this expression in \eqref{p4keisan}, we obtain
\[
p_4=a^{k(k+2)}p_1^{k+1}p_2^k\left\{-a^{k^2-k-1}X+p_1^{k^2-1}p_2^{k^2}(b-a)    \right\}
+a^{2k(k+1)}p_1^{k(k+1)}c^{k+1}\tilde{p}_3^k,
\]
which indicates
\[
\tilde{p}_4(0,b,c)=p_1^{k(k+1)}p_2^{k(k+1)}b\big|_{a=0}=c^{k(k+1)^2(k+2)}b \neq 0.
\]
Thus we have proved that ord$_a (p_4)=k(k+2)=\beta_4$.

In the case of $n = 5$,
we have from expression \eqref{eqp1},
\[
p_5=a^{k(k+1)(k+2)}\left[ \tilde{p}_4^{k+1}-a\tilde{p}_4^k\tilde{p}_3^{k+1} + c^{k+1} ( p_1p_2\tilde{p}_3)^{k(k+1)}\right].
\]
We have $\tilde{p}_5(0,b,c)=(a^{-k(k+1)(k+2)}p_5)\big|_{a=0} =c^{k(k+1)^3(k+2)}(b^{k+1}+c^{k+1}) \neq 0$.
Therefore we have proved that ord$_a(p_5)=k(k+1)(k+2)=\beta_5$.

Finally we prove the case of $n \ge 6$.
From the definition of $\beta_n$, we have
\[
\beta_{n}=(k+1)\beta_{n-1}=k\beta_{n-1}+(k+1)\beta_{n-2}=k(k+1)\sum_{j=0}^{n-2}\beta_j.
\]
Therefore we have from \eqref{eqp1} for $n\ge 6$ that
\begin{equation}
\tilde{p}_n=\tilde{p}_{n-1}^{k+1}-\tilde{p}_{n-1}^k\tilde{p}_{n-2}^{k+1} + c^{k+1} (\tilde{p}_1\tilde{p}_2...\tilde{p}_{n-2})^{k(k+1)}, \label{tildepnkeisan}
\end{equation}
which clearly indicates that $\tilde{p}_n$ is a polynomial. 
If we define $\displaystyle z'_n=\frac{\tilde{p}_n}{(\tilde{p}_{n-1}\tilde{p}_{n-2}...\tilde{p}_1)^k}$ and $z_n:=z'_n\big|_{(a=0,c=1)}$, we have $z_4=b$ and $z_5=(1+b^{k+1})/b^k$.
By shifting the subscript $n$ to $n+1$ in equation \eqref{tildepnkeisan}, and then by dividing both sides by 
$(\tilde{p}_n\tilde{p}_{n-1}\cdots \tilde{p}_1)^k$,
we have for $n \ge 5$ that
$z'_{n+1}=-z'_{n-1}+z'_n+c^{k+1}/(z_n')^k$.
By substituting $a=0$ and $c=1$ we have
\[
z_{n+1}=-z_{n-1}+z_n+\frac{1}{z_n^k}.
\]
This recurrence relation gives the same solution as \eqref{HVeq}
with initial conditions $z_3=0,z_4=b$. Therefore lemma \ref{lem1} tells us that $z_n$ is not identically zero.
We have proved $\tilde{p}_n(0,b,1) \neq 0$.
\qed
\begin{Lemma}\label{lem3}
The general term $x_n (n\ge 0)$ of the extended Hietarinta-Viallet mapping \eqref{HVeq}
for even $k\ge 2$ is expressed by polynomials $p'_n$'s as follows:
\begin{equation}\label{xpeq1}
x_n=\frac{p_n'p_{n-3}'}{c(p_{n-1}'p_{n-2}')^k}.
\end{equation}
Here we have defined formally as $p'_{-3}=p'_{-2}=p'_{-1}=1$.
\end{Lemma}
\Proof
We use
\[
x_n=\frac{p_n}{r_n}=\frac{p_n}{c(p_0p_1...p_{n-1})^k},
\]
and the relation \eqref{pnexp}.
Let us denote the exponent of $p_{n-j}'$ $(0 \le j \le n)$
in the numerator $p_n$ as $I_{n-j}$.
From lemma \ref{lem2}, we have $I_{n-j}=\beta_j$.
As for the denominator $c(p_0p_1...p_{n-1})^k$, let us denote the exponent of $p_{n-j}'$ as $J_{n-j}$. Then again from lemma \ref{lem2}, we have $J_{n-j}=k\sum_{i=0}^{j-1}\beta_i$.
For $j\ge 5$, we have
\[
J_{n-j}=k\left(1+(k+1)+\sum_{i=4}^{j-1}k(k+2)(k+1)^{i-4}\right)=k(k+2)(k+1)^{j-4}=\beta_j.
\]
Therefore
\begin{equation*}
\left\{J_{n-j} \right\}_{j=0}^{n}=\{\beta_0-1,\beta_1+k,\beta_2+k, \beta_3-1,\beta_4,\beta_5,...,\beta_n\}.
\end{equation*}
Thus the exponent of $p_{n-j}'$ in $x_n$ is obtained by
\[
\left\{ (I_{n-j}-J_{n-j})\right\}_{j=0}^n=\{1,-k,-k,1,0,0,\cdots,0\},
\]
which proves equation \eqref{xpeq1}.
\qed

\begin{Lemma}\label{lem5}
For every $n=0,1,2,\cdots$,
any pair from the three polynomials $\{p_n',p_{n+1}',p_{n+2}'\}$ is coprime.
\end{Lemma}
\Proof
By substituting \eqref{xpeq1} in the mapping \eqref{HVeq}, we obtain the following equation for $p_n'$, where we have taken formally 
$p_{-1}'=p_{-2}'=p_{-3}'=1$:
\begin{equation}\label{eq9}
p_{n+1}'=\frac{{p'}_{n-3}^{k+1}{p'}_n^{k+1}-p_{n-4}'{p'}_{n-1}^{k+1}{p'}_n^k+c^{k+1}{p'}_{n-2}^{k(k+1)}{p'}_{n-1}^{k(k+1)}}{{p'}_{n-3}^k {p'}_{n-2}^{k+1}}.
\end{equation}
The lemma is proved inductively.
First, $p_2',\,p_1',\,p_0'$ are coprime.
Let us suppose that $p_m',\,p_{m-1}',\,p_{m-2}'$ are coprime for every $2\le m\le n$ and prove the case of $m=n+1$.
We can prove the co-primeness of $p_{n+1}'$ and $p_n'$ as follows:
Let us suppose that they have a common factor $w$, then equation \eqref{eq9} tells us that either $p_{n-1}'$ or $p_{n-2}'$ should have the same factor $w$. However, both of these cases contradict the co-primeness of
$p'_n, p'_{n-1}, p'_{n-2}$.
In the same manner, suppose that $p_{n-1}'$ shares a common factor $w_2$
with $p_{n+1}'$, then either $p'_n$ or $p'_{n-3}$ should have $w_2$ as a factor, which again leads to a contradiction.
Therefore the lemma is true for $m=n+1$.
\qed

Note that we shall prove a stronger statement that `every pair of two polynomials in $\{p'_n\}$ are coprime for even $k$ (when $c=1$)' in the last section of this paper,
although lemma \ref{lem5} is strong enough for our purpose to obtain the algebraic entropy.
\begin{Theorem}
Let us denote the degrees by $d_n:=\deg x_n$ and $s_n:=\deg p'_n$.
Then we have the recurrence relation for $s_n$ as
\begin{equation}
s_n=k(s_{n-1}+s_{n-2})-s_{n-3}+1,
\label{evenrecs}
\end{equation}
for $n\ge 3$
with $s_0=1,s_1=k+1,s_2=(k+1)^2$.
The recurrence relation for $d_n$ is
\begin{equation}
d_{n}=(k+1)d_{n-1}-(k+1)d_{n-3}+d_{n-4},
\label{evenrecd}
\end{equation}
for $n\ge 4$ with $d_0=1$, $d_1=k+1$, $d_2=(k+1)^2$, $d_3=k(k+1)(k+2)+1$.
The relation between $d_n$ and $s_n$ for $n\ge 3$ is
\begin{equation}
d_n=s_n+s_{n-3}.
\end{equation}
\end{Theorem}
\Proof
For $n=0,1,2,3$ we can check by direct calculation.
By a definition of the degree of rational functions,
we have from lemma \ref{lem3} that $d_n=\max[s_n+s_{n-3},\, 1+k(s_{n-1}+s_{n-2})]$ for $n\ge 3$.
Here we have used lemmas \ref{lem4} and \ref{lem5} to ensure that the denominator and numerator of $x_n$ in lemma \ref{lem3}
do not share a factor. 
Moreover, we have in fact $s_n+s_{n-3}=1+k(s_{n-1}+s_{n-2})$, since we have taken a homogeneous coordinate, where $\deg p_n=\deg r_n$.
Thus the recurrence \eqref{evenrecs} and the relation $d_n=s_n+s_{n-3}$ are proved.
From these two equations, the recurrence \eqref{evenrecd} is immediate.
\qed

\begin{Corollary} \label{evenentropy}
For even $k\ge 2$, the algebraic entropy of the mapping \eqref{HVeq} is
\[
\lambda_k=\ln \left[ \frac{k+1+\sqrt{(k-1)(k+3)}}{2}  \right].
\]
\end{Corollary}
\Proof
Suppose that the degree of $x_n$ increases exponentially as $d_n \sim \lambda^n$. Then the value of $\lambda$ should be the largest real root of
\[
\lambda^4-(k+1)\lambda^3+(k+1)\lambda-1=(\lambda^2-1)(\lambda^2-(k+1)\lambda+1)=0,
\]
from the recurrence relation \eqref{evenrecd}.
\qed

Note that corollary \ref{evenentropy} is also true for $k=1$, since
in the case of $k=1$, the equation \eqref{HVeq} is integrable and has zero algebraic entropy.
Also note that every discussion in this subsection for even $k\ge 2$ is satisfied for $k=1$.
\subsection{The case of odd $\boldsymbol{k}\ge 3$}
Let us prove the proposition \ref{Prop1} for odd $k\ge 3$ in this subsection and obtain the algebraic entropy of \eqref{HVeq}.
Remember that we have defined the sequence $\beta_n$ $(n\ge 0)$ as
$\beta_0=1,\,\beta_1=\beta_2=0$ and
$\beta_n:=  k(\beta_{n-1}+\beta_{n-2})+(k+1)\beta_{n-3}$
for $n\ge 3$.
First let us prepare a simple lemma:
\begin{Lemma}\label{lembcd}
Let us define
\[
B_n^{(2)}:=k\beta_{n-1}+k(k+1)\sum_{j=0}^{n-3}\beta_j,\quad 
B_n^{(3)}:=k(k+1)\sum_{j=0}^{n-2}\beta_j.
\]
Then, for $n \ge 3$, we have
\[
\min \left[ \beta_n, B_n^{(2)}, B_n^{(3)} \right]\ge \beta_n.
\]
\end{Lemma}
\Proof
For $n=3,4,5$ we have
$\beta_3=k+1\ <\ B_3^{(2)}=B_3^{(3)}=k(k+1)$, and
$\beta_4=B_4^{(3)}=k(k+1)\ <\ B_4^{(2)}=2k(k+1)$, and
$\beta_5=B_5^{(2)}=k(k+1)^2\ <\ B_5^{(3)}=k(k+1)(k+2)$.
We obtain inductively that
\begin{align*}
&\beta_n<B_n^{(2)}=B_n^{(3)} \qquad (n\equiv 0\!\!\mod 3), \\
&\beta_n=B_n^{(3)}<B_n^{(2)}  \qquad (n\equiv 1\!\!\mod 3), \\
&\beta_n=B_n^{(2)}<B_n^{(3)}  \qquad (n\equiv 2\!\!\mod 3).
\end{align*}
\qed

\Proofprop1
In the case of $n=0,1,2$ the proposition is trivial.
In the case of $n=3$, we have $\beta_3=k+1$ and equation \eqref{p3form}, which does not depend on the parity of $k$.
Therefore the proposition is proved.
In the case of $n=4$, we have $\beta_4=k(k+1)$.
We follow the calculation of $p_4$ in the case of even $k$ in equation \eqref{p4keisan}.
Then we have
\begin{equation*}
p_4=a^{k(k+1)}p_1^{k+1}\left\{ c^{k+1}a^{k(k+1)}p_1^{k^2-1}(\tilde{p}_3)^k +Y   \right\},
\end{equation*}
where
\[
Y=p_2^k \left[-a^{k^2-1}X+p_1^{k^2-1}p_2^{k^2}(a^{k+1}-a^k b+2c^{k+1}) \right].
\]
Here we have used the same polynomial $X$ as in the case of even $k$.
Since $Y$ is not divisible by a factor $a$, we have ord$_a(p_4)=k(k+1)=\beta_4$ and the case of $n=4$ is proved.
Let us prove the case of $n \ge 5$ by induction.
Let us assume that ord$_{a} (p_m)=\beta_m$ (i.e., if we define $\tilde{p}_m=a^{-\beta_m}p_m$, $\tilde{p}_m$ is a polynomial which is not divisible by $a$.) for $m\le n-1$.
From equation \eqref{eqp2} (with a shift $n\to n-1$), we have 
\[
\mbox{ord}\, _{a} (p_n)\ge \min[ k\beta_{n-1}+k\beta_{n-2}+(k+1)\beta_{n-3}, B_n^{(2)}, B_n^{(3)}]\ge \beta_n,
\]
from
\begin{equation} \label{B2B3eq}
\mbox{ord}_{a} (p_{n-1}^k(p_{n-3}p_{n-4}\cdots p_0)^{k+1})=B_n^{(2)},\; \mbox{ord}_{a}(p_{n-2}p_{n-3}\cdots p_0)^{k+1}=B_n^{(3)},
\end{equation}
and from lemma \ref{lembcd}.
We have proved that $\tilde{p}_n=a^{-\beta_n}p_n$ is a polynomial in $a,b,c$. Our final task is to prove that $\tilde{p}_n(a=0,b,c)$ is non-zero as a rational function of $b,c$, which is equivalent to ord$_{a}(p_n)\le \beta_n$.
The rest of the proof is not essential to the discussion below, and therefore will be found in the appendix.
\qed
\hbreak
Let us recall the definition of $p'_n$ in equations \eqref{alpha} and \eqref{pdash}.
Lemma \ref{lem2} tells us that $p'_n$ is a polynomial.
We have a decomposition of $x_n$ into powers of $p'_n$:
\begin{Lemma}\label{lemxoddex}
Let us define a parameter $\mu_n$ as
$\mu_{3m}=1,\,\mu_{3m+1}=\mu_{3m+2}=-k$ for $m\in\mathbb{Z}$.
Then $x_n$ is factored as 
\[
x_n=c^{-1}\prod_{j=0}^n(p_{n-j}')^{\mu_j} \label{xexpodd}=\frac{p_n'}{c(p_{n-1}'p_{n-2}')^k}\frac{p_{n-3}'}{(p_{n-4}'p_{n-5}')^k}...
\]
\end{Lemma}
\Proof
From
\[
x_n=\frac{p_n}{r_n}=\frac{p_n}{c(p_{n-1}p_{n-2}...p_0)^k},
\]
and from $p_n=\prod_{j=0}^n (p_{n-j}')^{\beta_j}$ in lemma \ref{lem2}, we have
\[
x_n=c^{-1}\prod_{j=0}^n (p_{n-j}')^{\beta_j-k\sum_{i=0}^{j-1}\beta_i},
\]
where we suppose that if $j=0$ the term $\sum_{i=0}^{j-1}\beta_i$ is zero.
For small $j=0,1,2,3$ we have
\[
\beta_0=1,\ \beta_1-k\beta_0=\beta_2-k(\beta_1+\beta_0)=-k,\ \beta_3-k(\beta_2+\beta_1+\beta_0)=1.
\]
Therefore the first four terms of factorization of $x_n$ are
$p_n'$, $(p_{n-1}')^{-k}$, $(p_{n-2}')^{-k}$, $p_{n-3}'$.
We easily prove that the power of $p_{n-j}'$ is periodic with period $3$ for $j\ge 1$, since
\[
\beta_j-k\sum_{i=0}^{j-1}\beta_i=\beta_{j-3}-k\sum_{i=0}^{j-4}\beta_i,
\]
from the definition of $\beta_n$.
Therefore equation \eqref{xexpodd} is proved.
\qed

\begin{Proposition}\label{propirredodd}
The polynomial $p_n'$ is coprime with every $p_j'$ with $0\le j<n$.
\end{Proposition}
\Proof
Let us define an auxiliary polynomial $R_n:=p_n'p_{n-3}'\dots p_{n-3[n/3]}'$, where the symbol $[y]$ denotes the largest integer that does not exceed $y$.
Lemma \ref{lemxoddex} indicates that
\[
x_n=\frac{R_n}{c R_{n-1}^k R_{n-2}^k}.
\]
By substituting this $x_n$ in the mapping \eqref{HVeq},
we obtain
\begin{equation}\label{inteq1}
p_{n+1}'=-(p_n'p_{n-1}')^k+c^{k+1}{p'}_n^kR_{n-2}^{k^2-1}R_{n-3}^{k(k+1)}+c^{k+1}R_{n-1}^{k(k+1)}R_{n-2}^{k^2-1}.
\end{equation}
The co-primeness is satisfied for $n=0,1,2$.
Let us assume that the proposition is true up to $p_n'$
and prove the co-primeness of $p_{n+1}'$ with $p_m'$ $(m\le n)$.
It is enough to prove that $p_{n+1}'$ is coprime with $R_n,R_{n-1},R_{n-2}$.
First, $p'_{n+1}$ is coprime with $p'_n$ from \eqref{inteq1}, since, otherwise, $p'_n$ has a common factor with $R_{n-1}$ or $R_{n-2}$, which contradicts the induction hypothesis.
In the same manner we have that $p'_{n+1}$ should be coprime with $p'_{n-1}$. Here we have used $R_{n-1}=p'_{n-1}R_{n-4}$.
From the co-primeness of $p'_{n+1}$ with $p'_{n}$ and $p'_{n-1}$ (, which we have just proved), and from equation \eqref{inteq1}, we obtain the co-primeness of $p'_{n+1}$ with $R_{n-2}$. The rest of the proof: i.e., the co-primeness of $p_{n+1}$ with $R_n$ and $R_{n-1}$, can be found in the appendix.
\qed

\begin{Theorem}\label{oddtheorem}
Let us denote the degrees as $t_n:=\deg p_n'$ and $d_n:=\deg x_n$.
The recurrence relations for $d_n$ and $t_n$ is given as
\begin{equation}\label{oddrecurrence}
d_{n+1}=(k+1)d_n-kd_{n-2}\;\; (n\ge 2),
\end{equation}
and
\begin{equation}\label{oddrect}
t_{n+1}=(k+1)t_{n}-k t_{n-2} \;\;(n\ge 3),
\end{equation}
with $t_0=d_0=1,\,t_1=d_1=k+1,\,t_2=d_2=(k+1)^2$, $t_3=k(k+1)(k+2)$.
The relation between $d_n$ and $t_n$ is
\begin{equation}\label{odddt}
t_n=d_n-d_{n-3}\;\; (n\ge 3).
\end{equation}
\end{Theorem}
\Proof
From lemma \ref{lem4} and proposition \ref{propirredodd},
the denominator and the numerator of the term $x_n$ in \eqref{xexpodd} do not share a common factor. From the homogeneous coordinates and the initial condition $[a:b:c]$, the degree of the denominator and the numerator of \eqref{xexpodd} must be the same.
Therefore we obtain
\begin{align*}
d_n&=t_n+t_{n-3}+\cdots+t_{n-3[n/3]}\\
&=1+k\left(t_{n-1}+t_{n-4}+\cdots+t_{n-1-3[(n-1)/3]}\right)\\
&\;\;+k\left(t_{n-2}+t_{n-5}+\cdots+t_{n-2-3[(n-2)/3]}\right)\\
&=1+k(d_{n-1}+d_{n-2}).
\end{align*}
It is straightforward to prove the recurrences \eqref{oddrecurrence}, \eqref{oddrect} and \eqref{odddt}.
\qed

From the recurrence \eqref{oddrecurrence} for $d_n$ we can obtain the algebraic entropy of \eqref{HVeq}.
\begin{Corollary} \label{algodd}
For odd $k\ge 3$, the algebraic entropy of the mapping \eqref{HVeq} is
\[
\lambda_k=\ln \left[\frac{k+\sqrt{k(k+4)}}{2}\right].
\]
\end{Corollary}
\section{Irreducibility of polynomials $p'_n$ for even $k\ge 2$} 
Let us reconsider the extended Hietarinta-Viallet equation where $k$ is
an even integer:
\begin{equation}
x_{n+1}=-x_{n-1}+x_n+\frac{1}{x_n^k} \qquad (k=2,4,6,...).
\label{HVeq2}
\end{equation}
We prove the irreducibility theorem \ref{evenirredthm}, which is stronger than lemma \ref{lem5} on the co-primeness of three consecutive iterates.
We limit ourselves to the case of $c=1$, since this case is enough for our purpose of the irreducibility of $x_n$ as a rational function of initial variables $x_{-1}=b$ and $x_0=a$.
Let us reproduce the equation of $p'_n$ in \eqref{eq9} here for $c=1$:
\begin{equation}\label{eqq9}
p_{n+1}'=\frac{{p'}_{n-3}^{k+1}{p'}_n^{k+1}-p_{n-4}'{p'}_{n-1}^{k+1}{p'}_n^k+{p'}_{n-2}^{k(k+1)}{p'}_{n-1}^{k(k+1)}}{{p'}_{n-3}^k {p'}_{n-2}^{k+1}}.
\end{equation}
If we formally take $p_{-4}'=b$, $p_{-3}'=p_{-2}'=p_{-1}'=1$ and $p_0'=a$, then $p'_n\in\mathbb{Z}[a,b]$ and the rational functions $x_n=(p_n'p_{n-3}')/(p_{n-1}'p_{n-2}')^k$ $(n\ge -1)$ satisfy
the mapping \eqref{HVeq} with initial conditions $x_{-1}=b$ and $x_0=a$.
Let us recall the definition of $\beta_n$ for even $k$ in definition \ref{defbeta},
and redefine $p_n=\prod_{j=0}^n{p_j'}^{\beta_{n-j}}$. Then we reproduce equation \eqref{eqp1} as
\begin{equation}
p_{n+1}=p_n^k(p_n-p_{n-1}^{k+1})+\left(p_{n-1}p_{n-2}...p_1p_0\right)^{k(k+1)}  \quad (n \ge 1) \label{eqpp1}.
\end{equation}
\begin{Lemma} \label{lemma1} 
The polynomial $p_n'$ is not divisible by a factor `$b$' for $n\ge 0$.
\end{Lemma}
\Proof
Let us take $x_{-1}=b=0$ and evolve the mapping \eqref{HVeq}.
Then from lemma \ref{lem1} we have $x_n\neq 0$ as a function of $a$. Therefore $p'_n$ should be non-zero for $b=0$.
\qed

Next we introduce a gauge transformation.
\begin{Lemma}\label{gauge_trans}
Let us take arbitrary sequence $\{p_n^{(0)}\}$ that satisfies equation
\eqref{eqq9} for every $n$.
We introduce a sequence of `gauge' functions $\{u_n\}$ that satisfies
\begin{equation}\label{irr_rec}
u_nu_{n-3}=(u_{n-1}u_{n-2})^k,
\end{equation}
where we suppose $u_n\neq 0$ for every $n$.
Then a new sequence of functions $\{p_n^{(1)}\}$ defined by $p_n^{(1)}:=u_n p_n^{(0)}$ is also a solution of equation \eqref{eqq9}.
\end{Lemma}
\Proof
By substituting $p_n^{(1)}=u_np_n^{(0)}$ in equation \eqref{eqq9},
we easily obtain that all the following equalities should be satisfied, in order for $p_n^{(1)}$ to be a solution of \eqref{eqq9}: 
\[
u_{n+1}=\frac{u_{n-3}^{k+1}u_n^{k+1}}{u_{n-3}^ku_{n-2}^{k+1}}=\frac{u_{n-4}u_{n-1}^{k+1}u_n^{k}}{u_{n-3}^ku_{n-2}^{k+1}}
=\frac{u_{n-2}^{k(k+1)}u_{n-1}^{k(k+1)}}{u_{n-3}^ku_{n-2}^{k+1}}.
\]
They are easily obtained from the recurrence relation \eqref{irr_rec}.
\qed

\begin{Definition}
We define the polynomial $P_n\in \mathbb{Z}[a,b]$ as
\begin{equation} \label{largepdef}
P_n(a,b):= p_n',
\end{equation}
where the initial values of $p'_n$ in \eqref{eqq9} are
\[
p_{-4}'=b,\,p_{-3}'=1,\, p_{-2}'=1,\, p_{-1}'=1,\,p_0'=a.
\]
\end{Definition}
\begin{Proposition}\label{propm}
If we define the sequence of rational functions $p'_n$ ($n=1,2,3,\cdots$), from equation \eqref{eqq9} and the initial values
\[
p_{-4}'=b,\,p_{-3}'=\mu_3,\, p_{-2}'=\mu_2,\, p_{-1}'=\mu_1,\,p_0'=a,
\]
and denote them by $Q_n:=p'_n$.
Then it satisfies
\begin{equation}\label{pnform}
Q_n(a,b)=u_n(\mu_1,\mu_2,\mu_3) P_n \left( \frac{a \mu_3}{(\mu_1\mu_2)^k}, \frac{\mu_1b}{(\mu_2\mu_3)^k} \right),
\end{equation}
where the polynomial $P_n$ is defined in \eqref{largepdef},
and the extra factor $u_n(\mu_1,\mu_2,\mu_3)$ is defined from the recurrence relation \eqref{irr_rec} and from the initial variables
\[
u_{-4}=\frac{(\mu_2\mu_3)^k}{\mu_1},\quad u_{-3}=\mu_3,\quad u_{-2}=\mu_2.
\]
We have $Q_n \in\mathbb{Z}[a^{\pm},b^{\pm},(\mu_1)^{\pm},(\mu_2)^{\pm},(\mu_3)^{\pm}]$: i.e., $Q_n$ is a Laurent polynomial of the initial data.
\end{Proposition}
\Proof
We define the sequence $\{x_n\}$ from the initial values
$x_0= (a \mu_3)/(\mu_1\mu_2)^k$ and $x_{-1}= (\mu_1 b)/(\mu_2\mu_3)^k$,
and the mapping \eqref{HVeq}.
Let us define another sequence $y_n=\{(q_n' q_{n-3}')/(q_{n-1}' q_{n-2}')^k\}$, using a sequence $q'_n$ obtained from equation \eqref{eqq9} and the initial values
\[
q_{-4}'=\frac{\mu_1 b}{(\mu_2\mu_3)^k},\quad q_{-3}'=q_{-2}'=q_{-1}'=1,\quad q_0'=\frac{a \mu_3}{(\mu_1\mu_2)^k}.
\]
Then $x_n=y_n$ for $n\ge -1$.
Therefore we have that
\[
q_n'=P_n \left( \frac{a \mu_3}{(\mu_1\mu_2)^k}, \frac{\mu_1b}{(\mu_2\mu_3)^k} \right).
\]
From lemma \ref{gauge_trans}, the sequence of polynomials $r'_n:=u_n q'_n$ should satisfy the equation \eqref{eqq9},
with initial values
$r_{-4}'=b$, $r_{-3}'=\mu_3$, $r_{-2}'=\mu_2$, $r_{-1}'=\mu_1$, $r_0'=a$.
(Note that $u_{-1}=\mu_1$, $u_0=(\mu_1\mu_2)^k/\mu_3$.)
Therefore the sequence $\{Q_n(a,b)\}$ in this proposition \ref{propm} coincides with $\{r'_n\}$ for every $n\ge -1$.
Thus $Q_n$ should be given by equation \eqref{pnform}.
The Laurentness of $Q_n$ is obtained from the fact that $P_n$ is a polynomial and the fact that $u_n$ is a monomial of $\mu_i$ $(i=1,2,3)$.
\qed

\begin{Proposition} \label{propkiyaku}
The Laurent polynomial $Q_n=p'_n \in \Z[a^\pm,b^\pm,\mu_1^\pm,\mu_2^\pm,\mu_3^\pm]$ is irreducible.
\end{Proposition}
\Proof The case of $n \le 0$ is trivial.
For $n=1$, the polynomial $p_1'$ is linear with respect to the variable `$b$', and therefore is irreducible.
We use a lemma on the factorization of the terms of discrete systems in our previous paper \cite{dKdVSC2} (This lemma basically states that the irreducibility is preserved by a shift of the variables, except for some monomial factors. We have reproduced it in the appendix as lemma \ref{maselemma}).
Then we obtain the following factorization of $p'_2$:
\begin{equation} \label{p2factor}
p_2'=(p_1')^d h,
\end{equation}
where $d\in \Z$, $d\ge 0$, and $h$ is irreducible in $\Z [a^\pm,b^\pm,\mu_1^\pm,\mu_2^\pm,\mu_3^\pm]$. 
If we take special initial values $b=-1,a=\mu_1=\mu_2=\mu_3=1$, we have from direct computation that
$p_1'=3$, $p_2'=2\cdot 3^{k}+1 \equiv 1 \mod 3$.
Therefore we have $d=0$. Thus the Laurent polynomial $p'_2=h$ is irreducible.
%
Since we have
$p_3'\equiv 1$, $p_4' \equiv -1$, $p_5' \equiv 1$, $p_6' \equiv -1$ mod $3$,
we can repeat the preceding argument to prove that $p'_n$ is irreducible for $n\le 6$.

For $n=7$, we again use lemma \ref{maselemma} as in the appendix to obtain two types of factorizations
\begin{equation} \label{p7factor}
p_7'=(p_1')^{c_1}g_1=(p_2')^{c_2}(p_3')^{c_3}(p_4')^{c_4}(p_5')^{c_5}(p_6')^{c_6}g_2,
\end{equation}
where $c_j\in\mathbb{Z}$, $c_j\ge 0$ $(1\le j\le 6)$, and that $g_1,g_2$ are irreducible in the ring $\mathbb{Z} [a^\pm,b^\pm,\mu_1^\pm,\mu_2^\pm,\mu_3^\pm]$.
Let us prove that $c_j=0$ for all $1\le j\le 6$ by contradiction.
From the irreducibility of $p'_1$ and $g_1$, at most one of $c_2,\cdots, c_6$ can be non-zero.
Thus we have only two possibilities of factorization of $p'_7$:
(i) If $c_2=\cdots=c_6=0$, then $p'_7=u p'_j$ for $j\in\{1,2,3,4,5,6\}$ and for some unit $u$,
(ii) If $c_j\neq 0$ for only one $j\in\{2,3,4,5,6\}$, then $p_7'=u p'_1 p_j'$ for some unit $u_2$.
Note that a unit is equivalent to a monomial of $a,b,\mu_j$ ($j=1,2,3$).
Let us eliminate the case (i). From proposition \ref{propm}, we have
\[
p_7'=u_7(\mu_1,\mu_2,\mu_3) P_7\left( \frac{a \mu_3}{(\mu_1\mu_2)^k}, \frac{\mu_1b}{(\mu_2\mu_3)^k} \right).
\]
Therefore $\hat{u}=u_7/u$ should be a unit from the irreducibility of $p'_j$ ($1\le j\le 6$). We again use proposition \ref{propm} to have
\[
P_7\left( \frac{a \mu_3}{(\mu_1\mu_2)^k}, \frac{\mu_1b}{(\mu_2\mu_3)^k } \right)
=\hat{u}u_j P_j\left( \frac{a \mu_3}{(\mu_1\mu_2)^k}, \frac{\mu_1b}{(\mu_2\mu_3)^k} \right).
\]
Here
$\hat{u}u_j$ should be a monomial of
$(a \mu_3)/(\mu_1\mu_2)^k$, $(\mu_1 b)/(\mu_2\mu_3)^k$.
If we impose $\mu_1=\mu_2=\mu_3=1$, then $\hat{u} u_j$ is a monomial
of $a$ and $b$. However, from lemmas \ref{lem2} and \ref{lemma1},
$\hat{u}u_j$ does not have a factor `$a$' or `$b$', from which we conclude that
$\hat{u}u_j=\pm 1$.
Thus we have that $\deg P_7=\deg P_j$, which is a contradiction.
The case (ii) is eliminated thorough a similar discussion used for (i) from $\deg P_7>\deg P_j+\deg P_1$.
We have proved that $p'_7$ is irreducible.
Exactly the same discussion applies to the case of $n\ge 8$, so that we obtain the irreducibility of $p'_n$.
\qed

\begin{Theorem} \label{evenirredthm}
The polynomial $P_n(a,b)\in \mathbb{Z}[a,b]$ is irreducible for every $n\ge 1$, where
$P_n(a,b)=p'_n$ is the general iterate of equation \eqref{eqq9} with initial values $p_{-4}'=b,\,p_{-3}'=1,\, p_{-2}'=1,\, p_{-1}'=1,\,p_0'=a$.
\end{Theorem}
\Proof
From propositions \ref{propm} and \ref{propkiyaku}, we have
\[
p_n'=u_n(\mu_1,\mu_2,\mu_3)P_n\left( \frac{a \mu_3}{(\mu_1\mu_2)^k}, \frac{\mu_1b}{(\mu_2\mu_3)^k} \right),
\]
and that $p'_n$ is irreducible in the ring
of Laurent polynomials $R:=\mathbb{Z}[a^{\pm},b^{\pm},\mu_1^{\pm},\mu_2^{\pm},\mu_3^{\pm}]$.
From lemma \ref{lem2}, the polynomial $P_n(x,y)$ is in $\mathbb{Z}[x,y]$.
Let us suppose that we have a decomposition $P_n(x,y)=f(x,y)g(x,y)$
into a product of polynomials $f,g\in\mathbb{Z}[x,y]$.
Let us define
\[
X:=\frac{a \mu_3}{(\mu_1\mu_2)^k},\;\;Y:=\frac{\mu_1b}{(\mu_2\mu_3)^k}.
\]
From the irreducibility of $p'_n$ in $R$,
either $f(X,Y)$ or $g(X,Y)$ should be a unit in $R$.
We suppose without loss of generality that $f(X,Y)$ is a unit in $R$. Then only the following form is allowed for $f(X,Y)$:
$f(X,Y)=X^{\lambda_1} Y^{\lambda_2}$, where $\lambda_1,\lambda_2\in\mathbb{Z}$. Note that $X$ and $Y$ themselves are units in $R$.
However, since $P_n$ does not have $a$ or $b$ as a factor from lemmas \ref{lem2} and \ref{lemma1}, we have $\lambda_1=\lambda_2=0$.
Thus $f(X,Y)=1$.
\qed

\section{Concluding remarks and discussions} \label{discuss}
In this paper, we studied an extended version of the Hietarinta-Viallet equation with one parameter $k\ge 2$ at the exponent of the last term. In the case of $k=2$, the original Hietarinta-Viallet equation is recovered.
We rigorously obtained its algebraic entropy $\lambda_k$ for every $k\ge 1$, by constructing the recurrence relation for the degrees of the iterates deg ($x_n$).
The extended Hietarinta-Viallet mapping has a positive algebraic entropy and is thought to be non-integrable for every $k\ge 2$. However, the pattern of singularities depends on the parity of $k$. For even $k$, the mapping passes the singularity confinement (SC) test, while, for odd $k$, it does not pass the SC test.
In corollary \ref{evenentropy}, we have proved that
$\lambda_k=\ln \{(k+1+\sqrt{(k-1)(k+3)})/2\}$ for even $k=2,4,6,\cdots$ (and also for $k=1$).
Note that, in the case of $k=1$, the mapping is an integrable autonomous version of the discrete Painlev\'{e} I equation, and has zero algebraic entropy.
In corollary \ref{algodd}, we have shown that $\lambda_k=\ln \{(k+\sqrt{k(k+4)})/2\}$ for odd $k=3,5,7,\cdots$.
Confinement of the singularities indicates a smaller algebraic entropy resulting from cancellations of additional factors than the non-confining case.
In fact, we have the inequality $(k+1+\sqrt{(k-1)(k+3)})<(k+\sqrt{k(k+4)})$ for every $k\ge 1$.
We have made clear the difference between even $k$ and odd $k$ cases in terms of the algebraic entropy, although the mapping is considered to be non-integrable in both cases.

Our result for even $k$ agrees with the result in the paper \cite{MW},
in which the algebraic entropy $\lambda_k$ is conjectured using their full deautonomisation method.
In the paper \cite{MW}, it is mentioned that a non-autonomous mapping
$x_{n+1}=-x_{n-1}+x_n+(-1)^n/x_n^k$ $(k=3,5,7,\cdots)$
passes the SC test, and it is conjectured that the algebraic entropy of this mapping is equal to $\ln \{(k+1+\sqrt{(k-1)(k+3)})/2\}$, which agrees with our result for even $k$. We wish to improve our method to non-autonomous systems in future works.

Let us discuss the blowing-up methods.
The entropy of the original equation ($k=2$) is well-known to be obtained by constructing the space of initial conditions \cite{Takenawa,Takenawa2}.
Let $X$ be a rational surface constructed by blowing-up the domain $\mathbb{P}^1\times \mathbb{P}^1$ fourteen times at the singularities of the Hietarinta-Viallet equation.
Then the mapping is lifted to an automorphism over $X$. The surface $X$ is called the space of initial conditions of the mapping.
The same discussion should be possible for mappings with confined singularities.
It is an interesting problem to construct the space of initial conditions for the mapping \eqref{HVeq} with $k=4,6,8,\cdots$, by applying the method of blowing-ups to $\mathbb{P}^1\times \mathbb{P}^1$. It is not known how many times of blowing-ups we need to obtain the space.
Our conjecture is that the least number of blowing-ups needed to make the mapping birational is $6k+2$ for even $k=4,6,8,\cdots$.
Note that it agrees with the results for $k=1,2$.
For odd $k\ge 3$, we can prove that it is {\em not} possible to obtain the space of initial conditions for the mapping \eqref{HVeq}. According to the results of Diller and Favre \cite{DF}, the dynamical degree of an automorphism of a projective surface is either $1$, reciprocal quadratic or a Salem number. From our corollary \ref{algodd}, the dynamical degree of equation \eqref{HVeq} for odd $k$ is none of these three types.
Thus for odd $k$, we cannot expect too much to obtain the algebraic entropy from a geometric approach.
We also have a conjecture on the Dynkin diagram describing the action of the extended Hietarinta-Viallet equation on the Picard group of exceptional curves. We hope to present theses results in a rigorous manner in future works.

It is also interesting that nonlinear mapping \eqref{eqq9} has the Laurent property (i.e., every term of the equation is a Laurent polynomial of the initial variables) although it is not a multilinear type nor does it seem to have direct connection with the cluster algebras  \cite{FZ}, unlike the well-known equations such as the Hirota-Miwa equation. We aim to study the mapping \eqref{eqq9} in relation to the generalized versions of the cluster algebras.
Another future problem is to study discrete systems which are described as recurrence relations of more than order three.
Since the mapping \eqref{HVeq} is of order three, we can consider
this mapping over the projective space $\mathbb{P}^2$ or $\mathbb{P}^1\times \mathbb{P}^1$, whose geometric properties are fairly well-known.
However, for the mapping of higher order, geometric considerations such as the blowing-up method over $\mathbb{P}^m$ or $(\mathbb{P}^1)^m$ ($m\ge 3$) include quite sophisticated algebraic geometry.
Our method in this article avoids these difficulties, and therefore is expected to be applicable to wide class of mappings and is also useful in finding novel integrable and quasi-integrable discrete systems.

\section*{Acknowledgments}
The authors wish to thank Profs. R. Willox, B. Grammaticos and J. Mada for useful comments.
This work is partially supported by Grants-in-Aid for JSPS Fellows $25\cdot 3088$, and the Program for Leading Graduate Schools, MEXT, Japan.

\appendix
\section{Appendix}
\subsection{Latter half of the proof of proposition \ref{Prop1} for odd $k$}
To prove that $\tilde{p}_n(a=0,b,c)$ is not identically zero, it is enough to define $z_n:=\tilde{p}(a=0,b=1,c=1)$
and prove that $z_n \neq 0$.
If we define three auxiliary variables as
\begin{align*}
z_n^{(1)}&:=-z_{n-1}^kz_{n-2}^{k}z_{n-3}^{k+1},\ \ z_n^{(2)}:=z_{n-1}^k(z_{n-3}z_{n-4}...z_0)^{k(k+1)},\\
z_n^{(3)}&:=\left(z_{n-2}z_{n-3}...z_0\right)^{k(k+1)},
\end{align*}
 we have $z_0=1,\,z_1=1,\,z_2=1,\,z_3=-1,\,z_4=2$ and
\[
z_n=\left(z_n^{(1)}+a^{B_n^{(2)}-\beta_n}z_n^{(2)}+a^{B_n^{(3)}-\beta_n}z_n^{(3)}\right)\Big|_{a=0},
\]
from equation \eqref{eqp2} and \eqref{B2B3eq}.
Therefore we have from the proof of lemma \ref{lembcd},
\[
z_n=\left\{
\begin{array}{cl}
z_n^{(1)}&\quad (n\equiv 0\mod 3),\\
z_n^{(1)}+z_n^{(3)}&\quad (n\equiv 1\mod 3),\\
z_n^{(1)}+z_n^{(2)}&\quad (n\equiv 2\mod 3).
\end{array}
\right.
\]
These equations tell us inductively that
\begin{align*}
&\quad z_n < 0,\ z_n^{(1)}<0,\ z_n^{(2)}>0,\ z_n^{(3)}>0\;\; (n=3m),\\ 
&\quad z_n > 0,\ z_n^{(1)}>0,\ z_n^{(2)}<0,\ z_n^{(3)}>0\;\; (n=3m+1),\\ 
&\quad z_n > 0,\ z_n^{(1)}>0,\ z_n^{(2)}>0,\ z_n^{(3)}>0\;\; (n=3m+2),
\end{align*}
for $m\ge 1$.
(In the case of $m=1$, we have $z_3^{(1)}=-1$, $z_3^{(2)}=z_3^{(3)}=1$, $z_4^{(1)}=z_4^{(3)}=1$,
$z_4^{(2)}=(-1)^k=-1$, $z_5^{(1)}=z_5^{(2)}=2^k$, $z_5^{(3)}=1$.)
Therefore we have proved that $z_n \ne 0$.
%
%
\subsection{Latter half of the proof of proposition \ref{propirredodd} for odd $k$}
To prove the co-primeness of $p_{n+1}'$ with $R_{n}, R_{n-1}$,
we need the following lemma \ref{lemma9}.
\begin{Lemma} \label{lemma9}
For arbitrary integer $m$, we have
\begin{eqnarray}
&-&(p_n'p_{n-1}')^k+mc^{k+1}R_{n-1}^{k(k+1)}R_{n-2}^{k^2-1}\notag\\
&\equiv& {p'}_{n-1}^{k(k+1)}{p'}_{n-2}^{k^2-1}\left[-(p_{n-3}'p_{n-4}')^k+(m+1)c^{k+1}R_{n-4}^{k(k+1)}R_{n-5}^{k^2-1}   \right]  \label{eqq1}\\
&\mbox{mod}& R_{n-3},\notag
\end{eqnarray}
\begin{eqnarray}
&-&(p_{n-1}')^k+mc^{k+1}R_{n-2}^{k^2-1}R_{n-3}^{k(k+1)} \notag\\
&\equiv& {p'}_{n-2}^{k^2-1}{p'}_{n-3}^{k(k+1)}\big[-{p'}_{n-4}^k+(m+1)c^{k+1}R_{n-5}^{k^2-1}R_{n-6}^{k(k+1)}    \big] \mbox{mod} R_{n-4}. \label{eqq2}
\end{eqnarray}
\end{Lemma}
\noindent
\textbf{Proof of proposition \ref{propirredodd}}\;\;
From equations \eqref{inteq1} and \eqref{eqq1} with $m=1$,
we obtain the co-primeness of $p_{n+1}'$ and $p_{n-3}'$.
This is proved as follows: if we suppose that $p'_{n+1}$
and $p'_{n-3}$ have a common factor $w$, from the co-primeness of $p'_{n+1}$ and $p'_n$ which has already been proved, we conclude that
$p_{n-3}'$ and $R_{n-4}R_{n-5}$ should share a factor $w$.
This contradicts the induction hypothesis.
Next, substituting equation \eqref{eqq1} with $m=1$ and $m=2$ repeatedly in equation \eqref{inteq1} gives
\begin{align*}
&p'_{n+1}\equiv c^{k+1}{p'}_n^k R_{n-2}^{k^2-1} R_{n-3}^{k^2+k}+{p'}_{n-1}^{k^2+k}{p'}_{n-2}^{k^2-1}\big[-{p'}_{n-4}^{k^2+k} {p'}_{n-5}^{k^2-1} {p'}_{n-6}^k {p'}_{n-7}^k\\
&+3c^{k+1}R_{n-4}^{k^2+k}R_{n-5}^{k^2-1}\big] \mod R_{n-6}.
\end{align*}
Thus, if we suppose that $p'_{n+1}$ and $p'_{n-6}$ has a common factor $v$, then $p'_{n-6}$ should share the factor $v$ with the last term
\[
3c^{k+1}{p'}_{n-1}^{k^2+k}{p'}_{n-2}^{k^2-1}R_{n-4}^{k^2+k}R_{n-5}^{k^2-1},
\]
which contradicts the induction hypothesis.
By repeatedly using the equation \eqref{eqq1} for appropriate $m$,
we have inductively that $p'_{n+1}$ is coprime with $p_{n-3m}$, and therefore is coprime with $R_n$.
By repeating equation \eqref{eqq2}, we can prove that $p_{n+1}'$ is coprime with $R_{n-1}$, in a similar manner to the previous discussion.
Therefore the term $p_{n+1}'$ is coprime with $p_j'$ $(j \le n)$.
\qed

\noindent
\textbf{Proof of lemma \ref{lemma9}}\;\;
In the proof we rewrite $p'_n$ as $p_n$, since we have only $p'_n$ here.
By substituting equation \eqref{inteq1} ($n\to n-1$) into the left hand side of equation \eqref{eqq1}, we have
\begin{eqnarray*}
&&-p_n^k p_{n-1}^k+m c^{k+1} R_{n-1}^{k^2+k} R_{n-2}^{k^2-1}\\
&\equiv &p_{n-1}^{k^2+k} p_{n-2}^{k^2-1}\left(p_{n-2}+mc^{k+1} R_{n-4}^{k^2+k} R_{n-5}^{k^2-1}\right) \mod R_{n-3},
\end{eqnarray*}
since $k$ is odd.
From equation \eqref{inteq1} (with $n\to n-3$) we have
\begin{equation*}
p_{n-2}\equiv -p_{n-3}^k p_{n-4}^k+c^{k+1}p_{n-3}^k R_{n-5}^{k^2-1} R_{n-6}^{k^2+k} +c^{k+1} R_{n-4}^{k^2+k} R_{n-5}^{k^2-1}\mod R_{n-3}.
\end{equation*}
From these two equations we obtain equation \eqref{eqq1}.
The equation \eqref{eqq2} is proved in a similar manner by using
equation \eqref{inteq1}.

\subsection{A factorization lemma of Laurent polynomials in \cite{dKdVSC2}}
Let us suppose that we impose a transformation of variables (from $\boldsymbol{q}$ to $\boldsymbol{p}$) to an irreducible Laurent polynomial $g(\boldsymbol{q})$. If the change of variables are given as Laurent polynomials with several conditions, the following lemma \ref{maselemma} assures that, under the representation with a new variables $\boldsymbol{p}$, additional factors of the Laurent polynomial $g$ are limited to monomial factors of $\boldsymbol{p}$.
\begin{Lemma}[\cite{dKdVSC2}] \label{maselemma} 
Let $\boldsymbol{p}=\{p_1,p_2,\cdots,p_m\}$ and $\boldsymbol{q}=\{q_1,q_2,\cdots ,q_m\}$ be two sets of independent variables with the properties
\begin{equation*}
p_j\in \mathbb{Z}\left[ \boldsymbol{q}^{\pm}\right],\;\;\;
q_j\in \mathbb{Z}\left[\boldsymbol{p}^{\pm} \right],
\end{equation*}
and suppose that $q_j$ is irreducible in the ring $\mathbb{Z}[\boldsymbol{p}^{\pm}]$,
for $j=1,2,\cdots, m$. Here we have used an multi-index $\boldsymbol{p}^{\pm}$ to denote $p_1^{\pm}, p_2^{\pm},\cdots ,p_m^{\pm}$
and so on.
Let us take an irreducible Laurent polynomial
$f(\boldsymbol{p})\in \mathbb{Z}\left[ \boldsymbol{p}^{\pm}\right]$,
and another Laurent polynomial $g(\boldsymbol{q}) \in \mathbb{Z}\left[ \boldsymbol{q}^{\pm}\right]$,
which satisfies $f(\boldsymbol{p})=g(\boldsymbol{q})$.
In these settings, the function $g$ is decomposed as
\[
g(\boldsymbol{q})=p_1^{r_1}p_2^{r_2}\cdots p_m^{r_m}\cdot \tilde{g}(\boldsymbol{q}),
\]
where $r_1,r_2, \cdots, r_m\in\mathbb{Z}$ and $\tilde{g}(\boldsymbol{q})$ is irreducible in $\mathbb{Z} \left[\boldsymbol{q}^{\pm}\right]$.
\end{Lemma}
Let us first explain how to derive equation \eqref{p2factor}:
By a $+1$-shift of variables as $b\to\mu_3$, $\mu_3\to\mu_2$, $\mu_2\to\mu_1$, $\mu_1\to a$, $a\to p'_1$, we have
$p_1' \to p_2'$. (Note that $a=p'_0$.)
Thus we can take $p'_2=f(\boldsymbol{p})=g(\boldsymbol{q})$, where
$\boldsymbol{p}=\{\mu_3,\mu_2,\mu_1,a,p'_1\}$, $\boldsymbol{q}=\{b,\mu_3,\mu_2,\mu_1,a\}$. Here we define the functions $f,g$ as follows: the function $g$ is equal to $p'_2$ defined from the initial values $p'_{-4}\to b$, $p'_{-3}\to \mu_3$ ,$p'_{-2}\to \mu_2$, $p'_{-1}\to \mu_1$, $p'_0\to a$, and the function $f$ is equal to $p'_1$ defined from $p'_{-4}\to \mu_3$, $p'_{-3}\to \mu_2$, $p'_{-2}\to \mu_1$, $p'_{-1}\to a$, $p'_0\to p'_1$.
We have $g(\boldsymbol{q})=\mu_3^{d_3}\mu_2^{d_2}\mu_1^{d_1}a^{d_0}(p'_1)^{d}\cdot \hat{h}$, where $\hat{h}$ is irreducible in $\mathbb{Z}[\boldsymbol{q}^{\pm}]$ and each $d_i\in\mathbb{Z}$, $d\in\mathbb{Z}$.
Since $\mu_3,\mu_2,\mu_1,a$ are monomials,
$h:=\mu_3^{d_3}\mu_2^{d_2}\mu_1^{d_1}a^{d_0}\hat{h}$ is also irreducible in $\mathbb{Z}[\boldsymbol{q}^{\pm}]$.
From the Laurentness of $g(\boldsymbol{q})$, we have $d\ge 0$.
The factorization \eqref{p2factor} is proved.
Lastly let us prove equation \eqref{p7factor}.
The first factorization of $p'_7$ in \eqref{p7factor} is obtained by taking
\[
\boldsymbol{q}=\{b,\mu_3,\mu_2,\mu_1,a\},\;\;\boldsymbol{p}=\{\mu_3,\mu_2,\mu_1,a,p'_1\},
\]
in lemma \ref{maselemma}.
The second one in \eqref{p7factor} is by
\[
\boldsymbol{q}=\{b,\mu_3,\mu_2,\mu_1,a\},\;\;\boldsymbol{p}=\{p'_2,p'_3,p'_4,p'_5,p'_6\}.
\]

%
%


\begin{thebibliography}{99}

\bibitem{SC}
Grammaticos, B., Ramani, A. and Papageorgiou, V.,
Do integrable mappings have the Painlev\'e property?,
\textit{Phys. Rev. Lett.} \textbf{67}, 1825--1828 (1999).

\bibitem{Conte}
Conte, R.,
The Painlev\'{e} property. One Century Later,
\textit{CRM Series in Mathematical Physics}, Springer, New York (1999).

\bibitem{BV}
Bellon, M. P. and Viallet, C. M.,
Algebraic Entropy,
\textit{Comm. Math. Phys.} \textbf{204}, 425--437 (1999).

\bibitem{DF}
Diller, J. and Favre, C.,
Dynamics of bimeromorphic maps of surfaces,
\textit{Amer. J. Math.} \textbf{123}, 1135--1169 (2001).

\bibitem{HV}
Hietarinta, J. and Viallet, C.,
Singularity confinement and chaos in discrete systems,
\textit{Phys. Rev. Lett.} \textbf{81}, 325--328 (1998).

\bibitem{MW}
A. Ramani, B. Grammaticos, R. Willox, T. Mase, M. Kanki
The redemption of singularity confinement,
\textit{J. Phys. A: Math. Theor.}, \textbf{48}, 11FT02 (2015).

\bibitem{Takenawa}
Takenawa, T.,
A geometric approach to singularity confinement and algebraic entropy,
\textit{J. Phys. A: Math. Gen.} \textbf{34}, L95--L102 (2001).

\bibitem{Takenawa2}
Takenawa, T.,
Algebraic entropy and the space of initial values for discrete dynamical systems,
\textit{J. Phys. A: Math. Gen.} \textbf{34}, 10533--10545 (2001).

\bibitem{Viallet}
Viallet, C-M.,
On the algebraic structure of rational discrete dynamical systems,
\textit{J. Phys. A: Math. Theor.} \textbf{48}, 16FT01 (2015).

\bibitem{dKdVSC2}
Kanki, M., Mase, T., Mada, J. and Tokihiro, T.,
Irreducibility and co-primeness as an integrability criterion for discrete equations,
\textit{J. Phys. A: Math. Theor.} \textbf{47}, 465204 (15pp) (2014).

\bibitem{dtoda}
Kanki, M., Mada, J. and Tokihiro, T.,
Integrability criterion in terms of coprime property for the discrete Toda equation,
\textit{J. Math. Phys.} \textbf{56}, 022706 (22pp) (2015).

\bibitem{Ramani}
Ramani, A., Grammaticos, B., Satsuma, J. and Mimura, N.,
Linearizable QRT mappings,
\textit{J. Phys. A: Math. Theor.} \textbf{44}, 425201 (2011).

\bibitem{Tsuda}
Tsuda, T., Grammaticos, B., Ramani, A. and Takenawa, T.,
A Class of Integrable and Nonintegrable Mappings and their Dynamics,
\textit{Lett. Math. Phys.} \textbf{82}, 39--49 (2007).

\bibitem{FZ}
Fomin, S. and Zelevinsky, A.,
The Laurent phenomenon,
\textit{Adv. Appl. Math.} \textbf{28}, 119--144 (2002).

\end{thebibliography}
\end{document}